\documentstyle[11pt,epsf]{article}

\def\beq{\begin{eqnarray}}
\def\eeq{\end{eqnarray}}
\def\bea{\begin{eqnarray*}}
\def\eea{\end{eqnarray*}}

\def\centeron#1#2{{\setbox0=\hbox{#1}\setbox1=\hbox{#2}\ifdim
\wd1>\wd0\kern.5\wd1\kern-.5\wd0\fi
\copy0\kern-.5\wd0\kern-.5\wd1\copy1\ifdim\wd0>\wd1
\kern.5\wd0\kern-.5\wd1\fi}}
\def\ltap{\;\centeron{\raise.35ex\hbox{$<$}}{\lower.65ex\hbox{$\sim$}}\;}
\def\gtap{\;\centeron{\raise.35ex\hbox{$>$}}{\lower.65ex\hbox{$\sim$}}\;}

\def\dslash{\not{\hbox{\kern-2pt $\partial$}}}
\def\Dslash{\not{\hbox{\kern-4pt $D$}}}
\def\Oslash{\not{\hbox{\kern-4pt $O$}}}
\def\Qslash{\not{\hbox{\kern-4pt $Q$}}}
\def\pslash{\not{\hbox{\kern-2.3pt $p$}}}
\def\kslash{\not{\hbox{\kern-2.3pt $k$}}}
\def\lslash{\not{\hbox{\kern-2.3pt $l$}}}
\def\qslash{\not{\hbox{\kern-2.3pt $q$}}}
\def\epsilonslash{\not{\hbox{\kern-2.3pt $\epsilon$}}}

\newcommand{\newc}{\newcommand}
\newc{\qbar}{{\overline q}}
\newc{\Kahler}{K\"ahler }
\newc{\deltaGS}{\delta_{\rm GS}}

\begin{document}

\begin{titlepage}
\begin{flushright}
{\large hep-th/0210255 \\  SCIPP-2002/20
}
\end{flushright}

\vskip 1.2cm

\begin{center}

{\LARGE\bf String Theory, Unification and Supersymmetry}

\vskip 1.4cm

{\large M. Dine, }
\vskip 0.2cm
%{\it $^a$Stanford Linear Accelerator Center,
%     Stanford CA 94309} \\
{\it Santa Cruz Institute for Particle Physics,
     Santa Cruz CA 95064  } \\
%\vskip 0.2cm
%{\large M. Graesser} \\
%\vskip 0.2cm
%{\it California Institute of Technology, 452-48,
%Pasadena, CA 91125}

%{\it $^c$Physics Department,
%     University of California,
%     Santa Cruz CA 95064  }
%\vskip 0.4cm
%{\large S. Thomas***}
%\vskip 0.2cm
%{\it $^a$Stanford Linear Accelerator Center,
%     Stanford CA 94309} \\
%{\it Physics Department, Stanford University, Stanford, CA  } \\

\vskip 4pt

\vskip 1.5cm

\begin{abstract}

One cannot yet point to any firm string prediction.  While
many approximate string ground states are known with interesting
properties, we do not have any argument that one or another
describes what we observe around us, and for reasons which
appear fundamental we do not know how to
systematically determine even any rough quantitative properties.
I argue here that we should examine large classes of string ground
states, trying to determine whether features such as low energy
supersymmetry, the pattern of supersymmetry breaking, the presence
of axions, large dimensions, or others might be generic.

\end{abstract}

\end{center}

\vskip 1.0 cm

\end{titlepage}

%%%%%%%%%%%%%%%%%%%%%%%%%%%%%%%%%%%%%%%%%%%%%%%%%%%%%%%%%%%%%%
% You may repeat \author \address as often as necessary      %
%%%%%%%%%%%%%%%%%%%%%%%%%%%%%%%%%%%%%%%%%%%%%%%%%%%%%%%%%%%%%%

\section{Introduction}

We have spent about 17 years thinking seriously about
string phenomenology.  In one sense we have come quite far.  We
have seen that string theory can exhibit many of the intricate
properties of the world we see:  four dimensions, Standard Model gauge
groups, repetitive generations, and the like.  As described at
this meeting, we know of string ground states which, in many of
their features, look quite close to the Standard Model.

Despite this -- indeed, partly because we have so many options for
thinking about the connection between string theory and nature -- we are hard pressed to
name a string theory prediction.  Does string theory predict low
energy supersymmetry?  Does it predict large extra dimensions
with/without low energy supersymmetry?  Does it predict gauge
coupling unification?  Neutrino masses of the sort observed?  If we
are honest, we must say we don't know.

In fairness, there are good reasons for this.  It has been clear
almost from the beginning that the problem of determining the
ground state of string theory and of making calculations in this
state is almost certainly a strong coupling problem in a
non-supersymmetric state.\cite{dineseiberg}
Despite all of the progress in non-perturbative
string theory, we still have no tools with which to approach such
problems.

So it is worthwhile, at a meeting like this, to pause and to ask:
how might string theory make contact with nature?  One can imagine
suggest several possibilities:
\begin{itemize}
\item We might hope to ``solve" the theory
and calculate everything we might wish to know.
As I have indicated, this seems implausible, for the forseeable
future.
\item
We might conjecture that some string solution is the
relevant one to describing nature, and manage to calculate some
quantities.  This is the subject of many of the talks at this
meeting, and more generally of most of the work on string
phenomenology.
Whether successful or not, this is an important activity since it teaches
us much about the underlying theory.  Here the goal might be to
predict, say, some ratio of fermion masses, or perhaps the
existence of some additional massive gauge bosons at accessible
energies.
\item
Finally, we might try to focus on some questions which have a
generic character.  We might try to argue that string theory
predicts low energy supersymmetry, or that, given low energy
supersymmetry, some pattern of symmetry breaking (anomaly
mediation, gauge mediation, gaugino mediation, etc.) is a likely
outcome.  A strategy to approach such problems might be to examine
broad classes of classical string ground states, trying to
determine features which are, or are not, typical.
\end{itemize}

It is this last possibility which will be the focus of this talk.
There is, of course, no guarantee of success.  But the potential
payoff is enormous.  Today
we will ask:
\begin{itemize}
\item
Is it possible to establish that low energy supersymmetry is a
prediction of string theory?
\item
Assuming that low energy supersymmetry is a prediction of string
theory, what can we hope to say about the pattern of soft
breaking?
\item
If the underlying
structure is a string theory, what might a grand unified field
theory look like?  And related to this:  what are the distinctions
between string-theoretic and field-theoretic unification?
\end{itemize}

 Our success today will be small but real.  We
will be able to give  examples of phenomena which are {\it not}
characteristic of typical string ground states.  In addition,
we will at least
take steps towards {\it defining} the problem of determining if low
energy supersymmetry is a prediction of string theory.  We will
describe an attempt to show that non-supersymmetric ground states
do not make sense in string theory.  We will give some indications
of how, given low energy supersymmetry, predictions for the
pattern of soft breaking might emerge.  We will argue that string
theory suggests a specific approach to conventional GUT model
building.

In recent years, theoretical speculations about unification of
forces have become more expansive, including large or very large
extra dimensions, and localized gravity.  I won't talk about these
ideas here, but note that none of these
ideas can be explored without a theory like string theory, which
incorporates gravity and is finite (``complete" might be a better
expression).
I also won't talk about flavor, since it is harder to see how to
make generic statements (but I would love to).

In the next section, I discuss what might be called
``String inspired Grand Unification."  I give some reasons why
grand unification is interesting, even in string theory, and argue
that string theory suggests some rules for grand unified
model building.  In section 3, I discuss some attempts
to predict low energy supersymmetry from string theory.  In
section 4, I review an analysis that shows that anomaly mediation
is not a generic outcome of string theory.  In section 5, I
discuss aspects of moduli in string theory, and their implications
for phenomenology.  In particular, I explain why some possible
solutions of the cosmological moduli problem point to particular
mechanisms for supersymmetry breaking.

\section{String theoretic unification}

There has, through the years, been much criticism directed at
string theory and string theorists, to the effect that string
theory is somehow not a scientific theory.  In assessing these
criticisms, it is useful to contrast the situation with grand
unification.  In the case of grand unification, one has an
essentially infinite array of theories.  Prior to the resurgence
of string theory in the mid '80's, there were many ad hoc rules
for writing down such models.  To mention two, many models were
postulated with global continuous symmetries and with axions.
Many questions were raised about whether these two ideas made
sense in a theory with gravity.
String theory early on offered an answer to both questions:
\begin{itemize}
\item
In string theory there are no continuous global symmetries.\cite{banksdixon}
There are, on the
other hand, often discrete symmetries.\cite{gsw}   These symmetries are often,
and likely always, gauge symmetries.
\item
There are axions associated with spontaneously broken global symmetries
which are exactly conserved in perturbation theory, and broken by
non-perturbative effects.
\end{itemize}

\subsection{String Unification vs. Field Theory Unification}

Coupling unification is one of the few solid pieces of evidence in
favor of supersymmetry, as well as unification.  While
supersymmetric guts predict approximate unification of coupling
constants,
string theory offered from the beginning a way to think about coupling unification
different than that of field theory.  In string theory,
unification can occur even if there is no scale at which physics
can be described by an effective four dimensional field theory
with a unified gauge group.
For
that matter, there need not even be any scale at which one sees a
higher dimensional theory.  This is illustrated by Calabi-Yau
compactifications at a Gepner point.\cite{gepner}
One of the triumphs of weakly coupled heterotic strings was their
generic prediction of unification of couplings at weak coupling.
These theories also yielded another remarkable result:   they
could readily explain the splitting between doublets and triplets.
No symmetry explanation was required; this is an example of a
string miracle.\cite{gsw}
Yet these ideas were have phenomenological limitations.
\begin{itemize}
\item  In weakly coupled heterotic string theory, the unification
scale is essentially equal to the Planck scale.  This is not
the case in the strongly coupled limit,\cite{wittency,bdhw} but at strong coupling it is not clear that
unification is a robust prediction.  Similar statements apply to
other regimes of the moduli space.
\item
Neutrino masses provide further evidence that there is another
scale in nature well below the Planck scale.  While this could
be something like the scale of the strongly coupled
heterotic theory, it is interesting to
explore the possibility that these are connected with a purely
four-dimensional scale.
\item
Leptogenesis\cite{buchmuller} suggests the existence of a scale well below the
Planck scale.  Baryogenesis through coherent scalar field
oscillations might provide an alternative,\cite{ad} but the existence of
substantial neutrino masses makes the leptogensis scenario seem a
promising one.
\end{itemize}

\subsection{The Unification Scale as a Modulus}

Many papers have been written on the possibility of obtaining
conventional unification in string theory.  By this one
means finding states with
adjoints of a group like $SU(5)$ in some limit of string theory
(e.g. weakly coupled heterotic strings).  In order that there be
a separation of scales, it is generally necessary that the
adjoints be massless in some approximation.
But this is not all one needs to build
a successful unified model.  One needs, as well
\begin{itemize}
\item  A flat or nearly flat potential for the adjoint, in order
that this field can obtain a very large vev.
\item  No extra massless octets and triplets of $SU(3)\times
SU(2)$ in this direction.
\end{itemize}

Obtaining flat directions naturally in a supersymmetric field theory is easy.
Suppose one has a discrete
$R$-symmetry under which the superpotential transforms as
\beq
W \rightarrow \alpha W ~~~~~~\alpha=e^{2 \pi i \over N},
\eeq
while the adjoint, $A$, is invariant, $A \rightarrow A$.
Then
$W = A^n$
is forbidden, for all n.  Such discrete $R$ symmetries are common
in string theory.  We will exploit such
symmetries throughout our discussion.
One should keep in mind, though, that conventional
notions of naturalness are not always
applicable in string theory and flat directions often arise for
which there is no obvious field-theoretic explanation.

This simple model above has a serious difficulty.  While there
is a flat direction, there are also a massless octet and triplet
of $SU(3) \times SU(2)$ in the low energy theory, which completely
spoil the prediction of unification.

It is not easy to fix this in $SU(5)$ or $O(10)$ with only
discrete symmetries, even if one adds additional
fields.\footnote{In O(10), Hall and Raby
have written a model with continuous global R symmetries without
this problem; for future reference, this model contains
six adjoints, two
symmetric tensors and two spinor representations just in order to obtain
the first stage of symmetry breaking.\cite{hallraby}}
As we will see, however,
it is not difficult in the context of the models which Witten has
recently described
to attempt to address the doublet-triplet problem.\cite{wittengut}

Witten has proposed that one should understand
the lightness of Higgs doublets
by supposing that there is a discrete symmetry
which distinguishes doublets and triplets;\cite{wittengut} ideas along
these lines are implicit in earlier
work of
Barr.\cite{barr}
The basic model-building strategy can be summarized by taking the group to be
$SU(5) \times SU(5)$, with two pairs of bifundamentals, $\Phi_i$,
$\bar \Phi_i$, with expectation values:
\beq
 \Phi_1 = \bar \Phi_1 =
\left ( \matrix{v_1 &  &  &  &  \cr  &  v_1  & & &  \cr
 &  & v_1 & &  \cr   &  & & 0 &  \cr  &    & & & 0}
\right ) ~~~~~
\Phi_2 = \bar \Phi_2 =
 \left ( \matrix{0 &  &  &  &  \cr  &  0  & & &  \cr
  &  & 0 & &  \cr   &  & & v_2 &  \cr  &    & & & v_2}
 \right )
 \label{phivevs}
\eeq
%\vskip 8cm
One can imagine\cite{barbierihall} then taking the Higgs fields
to be a $5$ and $\bar 5$ of one or the other $SU(5)$, and
coupling them only to $\Phi_1$.

This structure of expectation values is natural if it preserves a
symmetry.  This symmetry must be a combination of an ordinary
discrete symmetry and a gauge symmetry, say in the first $SU(5)$:
\beq
g_1 = {\rm diag}
 \left ( \alpha^{-1} ,\alpha^{-1}, \alpha^{-1} ,\alpha^{{N+3 \over
 2}},
  \alpha^{{N+3 \over 2}} \right )
 \eeq
Then if the $\Phi$'s transform under a $Z_N$ as:
\beq \Phi_1 \rightarrow \alpha \Phi_1
~~~~~\bar \Phi_1 \rightarrow \alpha^{-1} \bar \Phi_1
~~~~~
\Phi_2 \rightarrow \alpha^{-{N+3 \over 2}} \Phi_2
~~~~~\bar \Phi_2 \rightarrow \alpha^{{N+3 \over 2}} \bar \Phi_2
\eeq
the symmetry is preserved by the expectation values, and this
structure is natural.

A superpotential which respects this symmetry has been discussed by
Barr:
\beq
W= M_{gut}(r_1 \Phi_1 \bar \Phi_1 + r_2 \Phi_2 \bar \Phi_2)
+ {1 \over M_{gut}}(a~ \bar \Phi_1 \bar \Phi_1 \Phi_1 \bar \Phi_1
+ b~  \bar \Phi_1 \bar \Phi_1 \Phi_2 \bar \Phi_2 \eeq
 $$~~~~~
+ c~  \bar \Phi_2 \bar \Phi_2 \Phi_2 \bar \Phi_ + \dots
+ d~ {\rm tr} (\bar \Phi_1 \Phi_1) {\rm tr} (\bar \Phi_1 \Phi_1)
+e ~{\rm tr} (\bar \Phi_1 \Phi_1) {\rm tr} (\bar \Phi_2 \Phi_2)$$
$$
+ f ~{\rm tr} (\bar \Phi_2 \Phi_2) {\rm tr} (\bar \Phi_2 \Phi_2)+ \dots
).
$$

In this model, the GUT scale is put in explicitly.  There are no
extra states beyond those of the MSSM below this scale.

\subsection{Turning the GUT scale into a modulus}

To turn the GUT scale into an exact modulus in these
theories is not difficult;\cite{dns} we
can simply add a discrete R symmetry as for the $SU(5)$ model.  To also give
masses to all fields is more challenging (but not nearly as challenging
as in conventional $SU(5)$ and $O(10)$ models).  We can obtain
approximate moduli which accomplish this without great difficulty.
Giving mass to all fields requires, at a minimum, three adjoints and two singlets.
The fields and their transformation laws under the symmetries are
indicated in Table \ref{tab:modelb}.

\begin{table}[ht]
\begin{center}
\begin{tabular}{cc} \hline \hline
Field           &  $G_h$                \\ \hline\hline
$h$             &  $(5,1,\alpha,\beta)$         \\
$\bar h$        &  $(\bar5,1,1,1)$      \\
$h^\prime$      &  $(1,5,\alpha^{-1},\beta)$            \\
$\bar h^\prime$ &  $(1,\bar5,1,1)$      \\  \hline
$\Phi_1$        &  $(5,\bar5,\alpha,1)$ \\
$\bar\Phi_1$    &  $(\bar5,5,\alpha^{-1},1)$    \\
$\Phi_2$        &  $(5,\bar5,\alpha^{(N-3)/2},1)$       \\
$\bar\Phi_2$    &  $(\bar5,5,\alpha^{(N+3)/2},1)$       \\ \hline
$A_1$        &  $(24,1,\alpha^{(N-5)/2},\beta)$ \\
$A_2$           &  $(24,1,\alpha^{(N+5)/2},\beta)$ \\
$A_3$           &  $(24,1,1,\beta)$ \\  \hline
$S$             &  $(1,1,1,\beta^{-1})$ \\
\hline\hline
\end{tabular}
\end{center}
\caption{Field content of model I.}
\label{tab:modelb}
\end{table}

The renormalizable terms in the superpotential permitted by the
symmetries are:
$$
W= \lambda_{12}\Phi_1 A_1 \bar \Phi_2 + \lambda_{21}\Phi_2 A_2 \bar \Phi_1
+
\lambda_{11}\Phi_1 A_3 \bar \Phi_1 + \lambda_{22}\Phi_2  \bar \Phi_2+
\eta_{12} S A_1 A_2$$
\beq
+\eta_{33} \lambda^\prime S^\prime {A_3^2}
 +
X_1 {\rm tr}(A_1) + X_2 {\rm tr}(A_2) + X_3{\rm tr}(B).
\eeq
\noindent
In determining whether or not there are exact or nearly exact flat
directions, it is necessary to look beyond the renormalizable
terms.  An analysis of these shows that:
\begin{itemize}
\item  There are exact flat directions with $\Phi_i$'s non-zero, or $S$
non-zero, but not both; there are approximate flat directions with
both non-zero.  Phenomenologically, this can easily be good
enough.
\item  There are no extra states below the GUT scale.
Three adjoints constitute the smallest representation which
can achieve this.
\item It is difficult to build models with exact flat
directions for both $S$ and $\Phi$ and this low energy particle
content.
\item  It is not difficult to build models with ``baryonic" flat
directions, with $\Phi_1,\Phi_2 \ne 0$, $\bar \Phi_1,\bar \Phi_2
=0$, but this leaves a set of light fields with the quantum numbers of a full
$SU(5)$ adjoint, and the gauge couplings become
strong near the unification scale.
\end{itemize}

To summarize:  we have proposed a set of rules for grand unified
model building.  It is possible
to build models which satisfy them, but the rules are
very restrictive.  This is an appealing feature of
this viewpoint.

\subsection{Distinctions Between String Theory and Field Theory
Unification}

We have given above a definition of a grand unified field
theory (within the framework of
a more fundamental theory like string theory):  a theory in which,
for a range of scales, the standard model group is unified into a
larger group, with a finite number of fields.  We can use the
phrase
``string unification" to refer to theories in which couplings are
unified, but there is no range of scales in which the gauge
interactions are unified with a finite number of fields.
It is natural to ask:  are there
qualitative differences between string and field theory
unification?  Witten\cite{wittengut} has pointed out two:
\begin{itemize}
\item  String unification typically leads to superheavy
fractionally charged particles.
\item  Discrete symmetries in string theory are typically subject
to anomaly constraints.  In weakly coupled heterotic strings,
there is only one modulus which can cancel anomalies, so
discrete anomalies must be universal.  M. Graesser and I, however, have recently
shown that in other limits of string/M theory, discrete anomalies are
cancelled in by several different
axion-like fields, so there are no generic constraints.\cite{mdmg}
\end{itemize}

In addition, we have seen that, if field theoretic unification
arises within the framework of string theory,
the GUT scale is likely to be a modulus.  This has implications
for cosmology.

\section{Predicting Low Energy Supersymmetry?}

It is often said that supersymmetry is an integral part of string
theory, as if this somehow implies that low energy supersymmetry
is a feature of the theory.  But
supersymmetry is a gauge symmetry.  Just as for
ordinary bosonic symmetries, if supersymmetry is badly broken,
there is no low energy remnant.
How might we argue that low energy supersymmetry is a prediction
of string theory?

Clearly we should first ask:  what would it mean to predict low energy
supersymmetry in string theory?  In practice, what we have all
understood by this, and what I will understand in what follows, is
that the ground state of string theory which describes
the world we observe lies on an approximate $N=1$ supersymmetric
moduli space.  More precisely, it lives on a moduli space which
has the property that in various asymptotic regions, the theory
becomes four dimensional with approximate $N=1$ supersymmetry.  For
example, for the heterotic string on a Calabi-Yau manifold, the region
where the dilaton is large, with other moduli fixed, is
an example; for the
strongly coupled theory, one must take a different
limit.  In both cases, the potential energy is generically
non-zero throughout the moduli space, but tends to zero in this
limit.  This hypothesis is consistent with experimental facts:
the gauge couplings, for example, are small (corresponding to a
large value of the dilaton, perhaps) and there is a large
hierarchy, presumably related to some quantity like
$e^{-{8 \pi^2 \over g^2}}$.

One of the most difficult aspects of string phenomenology lies in
understanding the smallness of the gauge coupling and the large
size of the hierarchy.  String theory does not possess a small
parameter, in the sense of a quantity which can be taken
arbitrarily small.  So it would seem that, if the theory describes
nature, it must be strongly coupled.  Various scenarios for
understanding how a theory which is strongly coupled might yet
produce a small gauge coupling, or how the theory might ultimately
be weakly coupled, have been put forward.
To date, no complete string
implementation of any of these has been exhibited (several models of these
phenomena were discussed in the parallel sessions\cite{parralel}).

One might even have wondered whether it made sense to speak
of such approximate moduli spaces.  Indeed, it has always
been distressing that one can give arguments that moduli spaces of string
vacua with more than four supersymmetries exist, not only
perturbatively but non-perturbatively as well, whereas states with
less supersymmetry hold a more questionable status.
Most of the recent progress in string theory has been based
on the study of states with a high degree of supersymmetry, and, incidentally
has provided further evidence that such states exist and make sense.  But
the developments in duality have also provided evidence that moduli
spaces with approximate N=1 supersymmetry exist.

Much less is known about approximate
moduli spaces without supersymmetry.  In weak coupling, they
often exhibit pathologies.   Typically there are tachyons in some
region of the moduli space.  They are also often subject to
catastrophic instabilities.\cite{wittenkk}

Apart from
simply ``solving" the theory, one might imagine arguing that
supersymmetric
approximate moduli spaces -- and local minima which might
sometimes appear in them -- enjoy some special status.  In this
way one might argue that low energy supersymmetry is an outcome of
string theory.

In order to accomplish this, one must argue that moduli spaces with more
supersymmetry are, despite their apparent consistency,
irrelevant to describing what we see in nature.  Such an argument
might involve the connectivity of the moduli space; more likely,
it will involve cosmological considerations.  These might be connected
with some of the deep issues which have been raised recently
concerning the number of states in De Sitter space, or they might
be associated with some very weak anthropic considerations.
One might hope
to argue that non-supersymmetric moduli spaces are somehow
inconsistent, or perhaps disconnected from the supersymmetric
ones.

This program also requires that non-supersymmetric moduli spaces are
somehow inconsistent.  We have alluded to some evidence above.
The tachyon problem, naively, seems quite serious.  For example,
in weakly coupled, closed string theories, even if there is a minimum for
the tachyon potential, for fixed string coupling, the energy of
this minimum (the value of the dilaton potential) will behave,
parameterically, as
\beq
V = -{1 \over g^2} M^4
\eeq
for some mass scale $M$.  So it would appear that the energy is
unbounded below\footnote{This point was mentioned to me a few years
ago by L. Susskind.}  However, the situation is not so clear.
We do not know how to describe these systems as Hamiltonian
systems.  So it is best to examine their behavior in a
cosmological setting.  The full set of dynamical equations are
best studied by first integrating out the tachyon, then performing
a Weyl rescaling of the fields to the Einstein frame.  In this
way, one sees that the system is simply driven to strong coupling,
where one loses control of any analysis\footnote{If it happens
that the tachyon potential is only stabilized quantum
mechanically, then the system is driven to ever lower energy at
weak coupling.  In some cases, however, one can show that the
tachyon is stabilized classically.\cite{dgg}}, so it is difficult
to give a decisive argument that tachyons in a moduli space are
problematic.
The problem of catastrophic vacuum decay
first arose many years ago in work of Witten,\cite{wittenkk} and
has been the subject of more recent
analyses.\cite{fabingerhorava,horowitzetal}  Still another
possibility is to look for non-perturbative inconsistencies.
Michael Graesser and I have spent some time looking for anomalies
in discrete symmetries.\cite{mdmg}  Because discrete symmetries are believed
to be gauge symmetries in string theory, such anomalies would
signal inconsistencies.  Previous searches have been limited
to supersymmetric models.  There is no simple argument that such
anomalies cannot arise in non-supersymmetric theories.  We have
examined a variety of models, including asymmetric orbifolds and
various brane constructions.
So far, however, this search has not
yielded any positive results.

So, while there is some evidence that generic non-supersymmetric
string theories suffer from a variety of difficulties, we don't
have a solid, compelling argument that non-supersymmetric theories
do not make sense.  I have not given up on the
possibility that other sorts of anomalies might lurk in the
non-supersymmetric (approximate) moduli spaces.  It would be a
triumph if string theory were to successfully predict (or not) low energy
supersymmetry before its discovery (or not).  If we fail to make
such a prediction, the discovery of supersymmetry (or not) should
give us significant insight into the theory.

\section{String Theory in a Supersymmetric World}

If supersymmetry is discovered, whether or not string theory has
succeeded in predicting it, we will enter a very exciting era.
It is often said that string theory requires a Planck scale
accelerator, and that this is impossible.  But in most conjectures
for supersymmetry breaking,
the 105 or or more soft breaking parameters are related to
{\it Planck scale physics.}  So there is potentially a huge amount
of information accessible to TeV scale accelerators.

\subsection{Generalities about Supersymmetry Breaking}

At first sight, sorting this out may seem a daunting task, no easier than
understanding the quark and lepton masses. But we expect that the
squark and slepton mass spectrum will exhibit striking
regularities, to account for the absence of flavor-changing
processes.  Most proposals to understand approximate flavor
conservation involve a high degree of {\it degeneracy} or {\it
alignment} among the squarks and sleptons.

Only a few proposals have been put forward through the years to
understand degeneracy.  This could, of course, be due to
our lack of inventiveness.  But it is interesting to review them and
ask how they might fit into string theory.
\begin{itemize}
\item  Dilaton domination:  If the dilaton $F$-term dominates susy
breaking, this gives a  degenerate spectrum classically,\cite{dilatondomination}
with degeneracy (optimistically) of
order $\alpha_{{gut}} \over \pi$.  This is the only proposed realization
of ``gravity mediation" in string theory.  One difficulty with
this idea is that it requires that a weak coupling approximation
be valid for the Kahler potential, which is difficult to understand.
\item  Gaugino domination:\cite{gauginodomination}  If
the gaugino masses are much larger than scalar
masses at the high scale, then one obtains approximate
degeneracy through renormalization group evolution.
\item  Brane World Susy Breaking (Anomaly Mediation,\cite{anomalymediation} Gaugino
Mediation\cite{gauginomediation}).  These hypotheses give a predictive form for spectrum
with a high degree of degeneracy.
\item  Non-abelian flavor symmetries: These give degeneracy, correlations
between soft breakings, quark and lepton masses and mixings.
\item  Gauge mediation predicts a high degree of degeneracy
between squarks, sleptons with same gauge quantum numbers.\cite{gaugemediation}
More
detailed predictions are possible if the dynamics of supersymmetry breaking are
known.
\end{itemize}

How might these emerge as predictions from string theory, or what
might they tell us about string theory?  Might any of these have a
generic quality, e.g. true of a large set of string
states?  We have indicated how dilaton domination might arise.
Discrete symmetries, both abelian and non-abelian, are common in string
theory.  As we will see later, gauge mediation is a natural possibility
to consider in string theory, and might be a plausible
outcome of one solution of the moduli problem.\cite{dineshirman}  I am not aware of any
compelling picture of how gaugino domination might arise in a
generic way.  In the next
subsection, we will explain why one item in this list, anomaly
mediation (and gaugino mediation) does not seem to arise in string
theory in any generic sense.

\subsection{Almost a prediction (``unprediction"?))}

The basic idea of brane world susy breaking
is to suppose that the standard model localized
on a brane, while susy is broken on another brane.  Locality,
it is argued, strongly
constrains the interactions between the fields on different
branes, and the form of supersymmetry breaking. This leads
to vanishing scalar masses at low order, the leading contributions
being certain ``anomaly mediated" ones.\cite{anomalymediation}
Alternatively, if there are bulk gauge fields, the leading
contributions come from the interactions of these
fields.\cite{gauginomediation}

More precisely, crucial to anomaly mediated
supersymmetry breaking is the assumption that the Kahler potential
takes a particular form, which has been dubbed ``sequestered."
This form would seem to follow from higher dimensional locality.
Because the hypothesis does not make reference to the strength of
the coupling, this is a question which one can study in a variety
of string and M theory setups in controlled approximations.  In
all of these cases, one finds that the Kahler potential does not
have the sequestered form.\cite{adgt}
As a result, there are typically tree level masses, so the anomalous
and/or gaugino contributions do not dominate.  In fact,
one has problems with flavor changing currents unless there
are additional flavor symmetries.  So anomaly mediation
is not generic to string theory.

This is not to say that anomaly mediation could not emerge
in string theory.   A plausible argument has been
given that the required sequestered Kahler
potential might arise in special cases.\cite{lutysundrum}

While we have shown how one might rule out one possible form of
supersymmetry breaking, and suggested how others might arise from
string theory, we are clearly a way from making a definitive
prediction.  But I believe this discussion suggests that there is
some real hope.
These different possibilities make distinct and in some cases very
dramatic predictions for accelerators.

Apart from making predictions, these sorts of ideas suggest how
data we can expect over the next decade could provide important
information about string theory.  Imagine that supersymmetry has
been discovered, and that we know something about the spectrum.
If the spectrum is gauge-mediated, this would suggest possible
mechanisms for fixing the moduli.  If it is like gravity
mediation, but with percentish deviations and dramatic
flavor violation, this would be
suggestive of dilaton dominance.

\section{The Cosmological Moduli Problem, The Strong CP Problem And Other Issues}

The cosmological moduli\cite{cosmoproblem}
problem is usually described by saying that, in string
theory (in this section, we will take low energy supersymmetry
as a given), one might expect moduli to have a potential of the form:
\beq
V= m_{3/2}^2 M^2 f({\phi \over M}).
\eeq
Here $M$ is typically thought of as the Planck scale (within an
order of magnitude or two).
The field $\phi$ then has a mass of order $m_{3/2}$, and starts to
oscillate when $H \sim m_{3/2}$.  At this time, this field carries
a fraction of order $1$ of the energy density.  Even if there is
radiation at this time, $\phi$ quickly comes to dominate the
energy density of the universe.
The lifetime of $\phi$ is expected to be long, of order
\beq
\Gamma = {1 \over 2 \pi} {m_{3/2}^3 \over M^2}
\eeq
or smaller.

This is a long time after conventional nucleosynthesis.  The decays of
the moduli lead to some reheating:
$T_R^4 = {\Gamma^2 M^2}.$
For $m_{3/2} \approx 1~{\rm TeV}$, this gives a reheating
temperature of order $10~ {\rm KeV}$.  Restarting nucleosynthesis
requires $10~ {\rm MeV}$, which requires that the mass of the
modulus be of order $100~ {\rm TeV}.$

This is troubling from the perspective of fine tuning.  But even
if we accept this, there is another difficulty.\cite{bdg}  In most
discussions of the moduli multiplet, one speaks as if there is one
scalar.  But, of course, there are two.  In general, one of these
is an ``axion."  The defining property of this field, in many cases, is that it is
periodic.  We can take this period, with some suitable
normalization of the field, to be $2 \pi$.
In most pictures of supersymmetry breaking in string theory, this
field is light.
Most focus on supersymmetry breaking is on the superpotential.
Holomorphy plus $2 \pi$ periodicity strongly restrict the form of
the moduli superpotential:
\beq
W = e^{-a {\cal{M}}} + e^{-b {\cal{ M}}} + \dots
\eeq
for some constants $a$ and $b$.
So the full, supergravity potential will be of order
$e^{-2a {\cal M}}$, but the leading terms which violate the Peccei-Quinn
symmetry
will be suppressed by a further exponential, $e^{-(b-a) {\cal M}}$
This means that the axion in any given multiplet is light compared
to the scalar.  Yet this axion suffers from the same alignment
problems as the scalar.  If {\it it} is to give sufficient
reheating, the scale of supersymmetry breaking must be {\it very}
large.

To be more quantitative, there are several models for
stabilization of moduli in string theory.  One of these,
which has been mentioned in several talks at this meeting,
is known as Kahler stabilization.\cite{kahlerstabilization}
This assumes that $W$ is
given by some weak coupling
form, but order one corrections to the Kahler potential are responsible
for stabilization of the moduli.  For example, if gaugino condensation
is the origin of $W$,
\beq
m_{3/2}=e^{-3S/b_o}M
\eeq
giving $e^{-S/b_o} = 10^{-5}$.
Corrections to $W$ from, e.g.,
\beq
<W_\alpha^2 W_\beta^2> \sim
e^{-6S/b_o}
\eeq
give an axion mass seven orders of magnitude smaller than
the scalar mass ($10$ KeV?).  This is enough to cause cosmological
troubles, and far too large to be the QCD axion.

%An alternative model is known as the racetrack.\cite{racetrack}
%In the racetrack
%picture, $a$ and $b$ are nearly equal, so the suppression is
%now of order a power of $N$, rather than an exponential in the
%inverse coupling..  In fact, one expects
%\beq a \sim 1/N, b\sim
%1/N,a-b\sim 1/N^2 \approx \alpha
%\eeq
%for some large integer $N$.  The axion mass itself is
%${\cal O}(\sqrt{\alpha}$ relative to the scalar mass.  This still
%requires that the susy breaking scale be of order $10^3 {\rm
%TeV}$!

This problem would be solved if there were big corrections to the
Kahler potential which broke the symmetry,
e.g.
\beq
\delta K = e^{-({\cal M}-{\cal M}^\dagger)}f({\cal M}+{\cal M}^\dagger)
\eeq
(these respect the discrete shift symmetry, but badly violate the
PQ symmetry).
But if such corrections exist in string theory,
it will not be possible for such axions to solve
the strong CP problem.
One can consider many variations on this.  E.g. discrete
symmetries could suppress the PQ-violating terms in the
superpotential, and the Kahler potential corrections might be
small, allowing a small mass.

More generally, this discussion raises questions about the
strong CP problem in string theory -- and generally within
supersymmetric theories.\cite{bdg}  E.g. it is usually said that there is an
upper limit on the axion decay constant of about $10^{11}$ GeV.
This assumes that the universe is radiation dominated, for
example, at the time of the QCD phase transition.  But
supersymmetry alone implies that the axion is accompanied by a
scalar modulus.  For a decay constant of order $10^{15}$ GeV, for
example, the
cosmological problems associated with the saxion are far more
serious than those associated with the axion.  Any sensible
cosmology must address these, before claiming any limit on the
axion.
Results depend on model and cosmological assumptions, but rather
generally the limits on the axion decay constant are significantly
relaxed by these considerations.  $10^{15}$ GeV seems a rather robust
upper limit.\cite{bdg}

\section{A Unified Picture}

In focusing on the sorts of general questions I have discussed
here, we can all develop our own speculations on what sorts of
predictions we might hope to extract from string theory.
I would propose one outline of a generic string phenomenology.  No
piece of this can be viewed as firmly established, but I hope I
have indicated how such a picture might reliably emerge, using
tools we already have at hand:
\begin{itemize}
\item  Low energy supersymmetry
\item  No very light ($=m_{\rm susy}$) moduli (uncharged under symmetries;
there
may be charged moduli).
\item  One light ($ m_{susy} \ll m \ll M$)  modulus, determines the value of the
gauge couplings (one for unification). Its value might be
determined as in racetrack models.
\item
Because there are no light neutral fields, susy breaking arises through gauge
mediation at low energies.
\end{itemize}

Such a picture would be highly predictive, but by itself it would not
predict everything we might ultimately want to know.
For example, the soft
breaking spectrum would probably depend on a small number of unknown
parameters.  But this picture would have other interesting features:
\begin{itemize}
\item  No cosmological moduli problem
\item  No axion to solve the strong CP problem.  But in gauge
mediation, it is not difficult to solve strong CP problem through, e.g.,
the Nelson-Barr mechanism.\cite{dkl,schmaltz}
\item  More speculatively:  such a picture might be compatible with ideas of Banks to
understand the cosmological constant and supersymmetry breaking in de
Sitter space.\cite{banksdesitter}
\end{itemize}

I think this is a picture which might be correct and which we might
establish in string theory.  I am
optimistic that investigations of the sort I have
outlined here will teach us lessons about
the theory, even if my own prejudices are incorrect.  I encourage
you to explore this and other points of view.


\begin{thebibliography}{0}

\bibitem{dineseiberg}
M. Dine and N. Seiberg, Phys.Lett. {\bf B162} (1985) 299.



\bibitem{banksdixon}
T. Banks, L.J. Dixon, D. Friedan and E. Martinec,
Nucl.Phys. {\bf B299} (1988)613.



\bibitem{gsw}
M.B. Green, J.H. Schwarz and E. Witten,
{\it Superstring Theory,} Cambridge University
Press, Cambridge (1987).

\bibitem{gepner}
D. Gepner, Phys.Lett. {\bf B199}(1987) 380.

\bibitem{wittency}
E. Witten,
 Nucl. Phys. {\bf B471} (1996) 135 [arXiv hep-th/9602070].


\bibitem{bdhw}
T. Banks and M. Dine,
Nucl.Phys. {\bf B479} (1996) 173-196
[arXiv hep-th/9605136].


%\bibitem{wittency}
%E. Witten,
% Nucl. Phys. {\bf B471} (1996) 135 [arXiv hep-th/9602070].


%\bibitem{bdhw}
%T. Banks and M. Dine,
%Nucl.Phys. {\bf B479} (1996) 173-196
%[arXiv hep-th/9605136].

\bibitem{buchmuller}
A pedagogical introduction with extensive references appears in
W. Buchmuller, arXiv
hep-ph/0204288.

\bibitem{ad}
M. Dine and I. Affleck, Nucl.
Phys. {\bf B249}, 361 (1985);M. Dine, L. Randall
and S. Thomas,
Nucl.Phys. {\bf B458} (1996), 291 [arXiv hep-ph/9507453].


%\bibitem{wittency}
%E. Witten,
% Nucl. Phys. {\bf B471} (1996) 135 [arXiv hep-th/9602070].


%\bibitem{bdhw}
%T. Banks and M. Dine,
%Nucl.Phys. {\bf B479} (1996) 173-196
%[arXiv hep-th/9605136].


\bibitem{hallraby}
L. Hall and S. Raby,
Phys.Rev. {\bf D51} (1995) 6524 [arXiv hep-ph/9501298]

\bibitem{wittengut}
E.~Witten,
%``Deconstruction, G(2) holonomy, and doublet-triplet splitting,''
arXiv:hep-ph/0201018.
%%CITATION = HEP-PH 0201018;%%

\bibitem{barr}
S.M. Barr,
Phys.Rev. {\bf D55} (1997) 6775 [arXiv hep-ph/9607359].


\bibitem{barbierihall}
R.~Barbieri, G.~R.~Dvali and A.~Strumia,
%``Strings versus supersymmetric GUTs: Can they be reconciled?,''
Phys.\ Lett.\ B {\bf 333}, 79 (1994)
[arXiv:hep-ph/9404278].
%%CITATION = HEP-PH 9404278;%%

\bibitem{dns}
M. Dine, Y. Nir and Y. Shadmi,
arXiv hep-ph/0206268.


\bibitem{mdmg}
M. Dine and M. Graesser, in preparation.

\bibitem{parralel}
See the talks by T. Dent and J. Giedt at this meeting:
T. Dent, arXiv hep-ph/0208164; J. Giedt, arXiv hep-ph/0208004.s

\bibitem{wittenkk}
E. Witten,
 Nucl.Phys. {\bf B195} (1982) 481.

\bibitem{dgg}
M. Dine, E. Gorbatov and M. Graesser, in preparation.

\bibitem{fabingerhorava}
M. Fabinger and P. Horava,
Nucl.Phys. {\bf B580} (2000) 243 [arXive hep-th/0002073].

\bibitem{horowitzetal}
G.T. Horowitz and K. Maeda,
arXiv hep-th/0207270.

\bibitem{dilatondomination}
V. Kaplunovsky and J. Louis,
Phys. Lett. B {\bf 306} (1993) 269 [arXiv hep-th/9303040];
L. E. Ibanez and D. Lust, Nucl. Phys. B {\bf 382} (1992) 305.

\bibitem{gauginodomination}
M. Dine, A. Kagan and S. Samuel,
Phys.Lett. {\bf B243} (1990) 250.

\bibitem{anomalymediation}
L. Randall and R. Sundrum,
Nucl. Phys. B {\bf 557} (1999) 79 [arXiv hep-th/9810155];
G. Giudice, M.A. Luty, H. Murayama and R. Rattazzi
J. High Energy Phys. {\bf 9812} (1998) 027 [arXiv  hep-ph/9810442].

\bibitem{gauginomediation}
M. Schmaltz and W. Skiba,
+Phys. Rev. D {\bf 62} (2000) 095005 [arXiv  hep-ph/0001172];
D. Kaplan and G. Kribs,
J. High Energy Phys. {\bf 0009} (2000) 048 [arXiv hep-ph/0009195].

\bibitem{gaugemediation}
M. Dine and A. Nelson,
Phys.Rev. {\bf D48} (1993) 1277 [arXiv
hep-ph/9303230]; M. Dine, A.E. Nelson and Y. Shirman,
Phys.Rev. {\bf D51} (1995) 1362 [arXiv hep-ph/9408384];
M. Dine, A.E. Nelson, Y. Nir and Y. Shirman,
Phys.Rev. {\bf D53} (1996) 2568 [arXiv hep-ph/9507378];
G.F. Giudice and R. Rattazzi, Phys.Rept. {\bf 322} (1999) 419,
[arXiv hep-ph/9801271].



\bibitem{dineshirman}
M. Dine and Y. Shirman,
Phys.Rev. {\bf D63} (2001) 046005
[arXiv hep-th/9906246]; M. Dine,
Phys.Lett. {\bf B482} (2000) 213 [arXiv
hep-th/0002047].




\bibitem{adgt}
A. Anisimov, M. Dine, M. Graesser and S. Thomas,
arXiv hep-th/0201256 and
Phys.Rev. {\bf D65} (2002) 105011 [arXiv hep-th/0111235].

\bibitem{lutysundrum}
M.A. Luty and R. Sundrum, Phys.Rev. {\bf D65} (2002) 066004
[arXiv hep-th/0105137].

\bibitem{cosmoproblem}
T. Banks, D.B. Kaplan and A.E. Nelson,
Phys.Rev. {\bf D49} (1994) 779 [arXiv
hep-ph/9308292].

\bibitem{kahlerstabilization}
T. Banks and M. Dine,  Phys. Rev. {\bf D50} (1994) 7454,
arXiv hep-th/9406132.

%\bibitem{racetrack}
%N.V. Krasnikov, Phys. Lett. {\bf B193} (1987) 37;
%%%CITATION = PHLTA,B193,37;%%
%%\href{\wwwspires?j=PHLTA%2cB193%2c37}{SPIRES}
%L.J. Dixon, ``Supersymmetry Breaking in String Theory,"
%in {\it The Rice Meeting:  Proceedings}, B. Bonner and
%H. Miettinen, eds., World Scientific (Singapore) 1990;
%see \cite{dineshirman} for further discussion and references.



\bibitem{bdg}
T. Banks, M. Dine and M. Graesser in preparation

\bibitem{dkl}
M. Dine, R.G. Leigh and A. Kagan,
Phys.Rev. {\bf D48} (1993) 2214 [arXiv hep-ph/9303296].

\bibitem{schmaltz}
G. Hiller and M. Schmaltz,
Phys.Lett. {\bf B514} (2001) 263 [arXiv
hep-ph/0105254].

\bibitem{banksdesitter}
T. Banks, arXiv hep-th/0007146.

\end{thebibliography}
\end{document}